\documentstyle[preprint,aps,eqsecnum]{revtex}

\def\beq#1{\begin{equation} \label{#1}}
\def\eeq{\end{equation}}
\def\bra#1{\left\langle #1\right\vert}
\def\ket#1{\left\vert #1\right\rangle}

\def\PRL{ Phys. Rev. Lett.}

\begin{document}
{
\tighten

\title{Soft FSI Systematics for charmless strange $B^\pm$ Decays }

\author{Harry J. Lipkin\,$^{a,b}$}

\address{ \vbox{\vskip 0.truecm}
  $^a\;$Department of Particle Physics \\
  Weizmann Institute of Science, Rehovot 76100, Israel \\
\vbox{\vskip 0.truecm}
$^b\;$School of Physics and Astronomy \\
Raymond and Beverly Sackler Faculty of Exact Sciences \\
Tel Aviv University, Tel Aviv, Israel}

\maketitle

\begin{abstract}

New results going beyond those obtained from isospin and flavor symmetry
and subject to clear experimental tests are obtained for effects of FSI in
$B^\pm$ decays to final states containing neutral flavor-mixed mesons like
$\omega$, $\phi$, $\eta$ and $\eta'$. The most general strong-interaction
diagrams containing arbitrary numbers of quarks and gluons are included with
the assumptions that any $q \bar q$ pair created by gluons must be a
flavor singlet, and that there are no hairpin diagrams in which a final meson
contains a $q \bar q$ pair from the same gluon vertex. The smallness of
$K^- \eta$ suggests that it might have a large CP violation.
A sum rule is derived to test whether the large $K^- \eta'$ requires the
addition of an additional glueball or charm admixture.
Further analysis from $D_s$ decay systematics supports this picture of FSI and
raises questions about charm admixture in the $\eta'$.
\end{abstract}
} 

The successful treatment of strong final state interactions involving neutral
flavor-mixed mesons has a long history of successes going beyond isospin and
flavor SU(3) symmetry back to the
Alexander-Frankfurt-Harari-Iizuka-Levin-Okubo-Rosner-Scheck-Veneziano-Zweig
rule\cite{ALS,LevFran,HarDD,Venez}, often abbreviated A...Z or OZI. Its first
controversial prediction\cite{ALS}

\noindent $\sigma(K^-p \rightarrow \Lambda \rho^o)=
\sigma(K^-p \rightarrow \Lambda \omega)$ related final states in completely
different isospin and flavor-SU(3)
multiplets which would a priori be expected to have completely different final
state interactions. The experimental confirmation of this prediction\cite{ZGS}
indicates the existence of some dynamical symmetry that goes beyond isospin
and flavor SU(3).

Our purpose is to identify this symmetry and develop its
use to extend the standard isospin\cite{PBPENG} and SU(3)\cite{ROSGRO,etap}
treatments of $B$ decays to include flavor-mixed final states containing
$\omega$, $\phi$, $\eta$ and $\eta'$ mesons not easily treated in SU(3).
We also apply our new symmetry to otherwise
unexplained $D_s$ decay systematics\cite{PHAWAII,CLEODS}.

A recent application of the OZI or A...Z rule to B decay predicts
that\cite{PENGRHO}
$$ BR (B^\pm \rightarrow K^\pm \omega) = BR (B^\pm \rightarrow K^\pm \rho^o)
\eqno(1a)                                          $$
A similar approach also including broken SU(3) symmetry gives the prediction
$$ \tilde \Gamma(B^\pm \rightarrow K^\pm \phi)
\leq  \tilde \Gamma(B^\pm \rightarrow K^o \rho^\pm)
\eqno(1b)                                          $$
where $\tilde \Gamma$ denotes the theoretical partial width without phase space
corrections, and the equality holds under the assumption of SU(3) flavor
symmetry. The previous derivation justified
the relation between different SU(3) amplitudes by a hand-waving asymptotic
freedom argument enabling the final mesons to escape without final state
interactions.

We show here that these relations hold even in the presence of strong final
state rescattering via all possible diagrams involving quarks and gluons
in which all quark-antiquark pairs created by gluons are flavor singlets and
A...Z-forbidden disconnected ``hairpin diagrams" are excluded.
This is consistent with a large variety of experimental results and theoretical
analyses for strong interaction three-point and four-point
functions\cite{ALS,Venez,Exter} expressed by the duality
diagrams\cite{HarDD} of old-fashioned Regge phenomenology or the more modern
planar quark diagrams in large $N_c$ QCD\cite{PAQMREV,Precoc}
in which the leading Regge t-channel exchanges are
dual to s-channel resonances\cite{HJLANN,JENSEN}. This clear
assumption leads to predictive power; e.g. eqs. (1) which can be tested with
future experimental data.

We begin by exploiting known\cite{HJLCharm} flavor-topology\cite{CLOLIP12}
characteristics of
charmless strange $B^\pm$ decays. The final states considered for $B^-$ decay
all have the
quark composition $s \bar u q \bar q$ where $q \bar q$ denotes a pair of the
same flavor which can be $u \bar u$ , $d \bar d$ or $s \bar s$, and we do not
consider the possibility of charm admixture in the final state. The $q \bar q$
pair may come from a very complicated diagram involving many quarks and gluons.
But there are only two possibilities for its origin illustrated by figs. 1 and 2
of ref.\cite{HJLCharm}: (1) It is created by gluons
and must therefore be a flavor singlet denoted by $(q \bar q)_1$; (2) The quark
is a $u$ quark from the weak vertex and the pair can only be $u \bar u$. The
decays are described by
three parameters, an $s \bar u (q \bar q)_1$ amplitude, a $K^- u \bar u$
amplitude and a relative phase. The one relation obtainable between the
decays to four final states is shown below to be the sum rule:
$$ \tilde \Gamma(B^\pm \rightarrow K^\pm \eta') + \tilde \Gamma(B^\pm
\rightarrow K^\pm \eta) \leq  \tilde \Gamma(B^\pm \rightarrow K^\pm \pi^o)
+  \tilde \Gamma(B^\pm \rightarrow \tilde K^o \pi^\pm)
\eqno(2)                                          $$
where $\tilde K^o$ denotes $ K^o$ for the $B^+$ decay and $\bar K^o$ for the
$B^-$ decay, the equality holds in the flavor-SU(3) limit, the direction of the
inequality follows from the assumption that SU(3) symmetry breaking will
suppress the $s \bar s$ contribution to the singlet $(q \bar q)_1$ and the
result holds for any $\eta - \eta'$ mixing angle.
These sum rules are of particular interest because of the large experimentally
observed branching ratio\cite{CLEO} for $B^\pm \rightarrow K^\pm \eta'$. A
violation favoring $B^\pm \rightarrow K^\pm \eta'$ can
provide convincing evidence for an additional contribution\cite{PengSU3} like a
glueball, charm admixture\cite{Hareta,FRIJACK,ATSON} in the $\eta'$ wave
function or an A...Z-violating hairpin diagram. Present data indicate a
violation of between one and two standard deviations. If this violation is
confirmed by better data, the best candidate is the charm admixture originally
suggested by Harari\cite{Hareta} which still remains the only simple
explanation for the anomalously large branching ratio for the apparently
A...Z-violating cascade decay $\psi' \rightarrow \eta \psi$ and the failure to
observe the analogous cascade decays\cite{PDG}
$\Upsilon(nS) \rightarrow \eta \Upsilon(1S)$. An A...Z-violating gluonic
hairpin would contribute to all these cascades on the same footing.

In the kaon-vector (KV) system the ideal mixing of the $\omega-\phi$ system
simplifies the treatment to give the equality (1a) and the simplified sum
rule (1b).

In these $B^\pm$ decays all amplitudes arising from
the $b \rightarrow u \bar u s$ transition depend only upon a single sum of the
color-favored and color-suppressed tree contributions. This simplification
provides predictive power and allows crucial tests of the basic assumptions but
does not arise in the neutral decays. Thus amplitudes derived here
for charged decays are not simply related by isospin to amplitudes for neutral
decays.

We now examine the dependence of these amplitudes on CKM matrix elements.
The two $b$ quark decay topologies contributing to these
decays, $b \rightarrow c \bar c s$ and $b \rightarrow u \bar u s$, depend upon
two different products of CKM matrix elements. Their interference can give rise
to direct CP violation.

Assuming SU(3) flavor symmetry, the A...Z rule and a standard pseudoscalar
mixing\cite{Bramon,PengSU3},
$$ \ket{\eta} = {1\over{\sqrt 3}}\cdot(\ket{P_u} + \ket{P_d} -\ket{P_s});
 ~ ~ ~ \ket{\eta'} = {1\over{\sqrt 6}}\cdot(\ket{P_u} + \ket{P_d} + 2\ket{P_s})
 \eqno (3a)$$
$$ \ket{\pi^o} = {1\over{\sqrt 2}}\cdot(\ket{P_u} - \ket{P_d})  \eqno (3b)$$
gives three types of contributions for the $B^-$ decay into a kaon
and a pseudoscalar meson.
$$ B^- \rightarrow \bar u c \bar c s \rightarrow u \bar  s + (q \bar q)_1
\rightarrow \ket{R} \eqno (4a)$$
$$ B^- \rightarrow \bar u u \bar u s \rightarrow \ket{K^- P_u}
\eqno (4b)$$
$$ B^- \rightarrow \bar u u \bar u s \rightarrow u \bar  s + (q \bar q)_1
\rightarrow \ket{R}
\eqno (4c)$$
where the state $\ket{R}$ is defined as
$$\ket{R} \equiv {1\over{\sqrt 3}}\cdot(\ket{K^- P_u} + \ket{K^- P_s}
+ \ket{\bar K^o\pi^-})
\eqno (5a)$$
and $P_u$, $P_d$ and $P_s$ denote the $u \bar u$, $d \bar d$ and $s \bar s$
components in the $\pi^o$, $\eta$ and $\eta'$ pseudoscalar mesons.
This gives the result
 $$\ket{R} = {1\over{\sqrt 6}}\cdot \ket{K^- \pi^o} +
{1\over{\sqrt 2}}\cdot \ket{K^- \eta' } +
{\xi\over{\sqrt 2}}\cdot \ket{K^- \eta } +
{1\over{\sqrt 3}}\cdot\ket{\bar K^o\pi^-}
 \eqno (5b)$$
where $\xi$ is a small parameter to introduce a $K \eta$ contribution which
vanishes in the SU(3) symmetry limit with the particular mixing\cite{Bramon}
angle (3) as a result of a cancellation between the contributions from the
$P_u$ and $P_s$ components in the $\eta$ wave function. A small but finite
value of $\xi$ is suggested for realistic models by the $K \eta$ suppression
observed in other experimental transitions\cite{HJLCharm,PKETA} like decays of
strong $K^*$ resonances known to proceed via an even parity $u \bar  s$
+ singlet state. The possibility of a relatively large $CP$ violation in a
small
$K \eta$ branching ratio is discussed below.

The description (5) of the final state also expresses the contribution
both of the penguin diagram\cite{PKETA,PENGRHO} and of other diagrams
proportional to the $b \rightarrow c \bar c s$ vertex where
the $c \bar c $ pair is annihilated via a final state interaction.

The $B^-$ transition into any kaon-pseudoscalar state $\ket{f}$ is then
$$ \bra{f} T \ket{B^-} = A \langle f \ket{R} + B \langle f \ket{K^- P_u}  +
C \sqrt {3} \cdot \langle f \ket{R} \cdot \langle R \ket{K^- P_u}
 \eqno (6a)$$
$$ \bra{f} T \ket{B^-} = (A + C )\langle f \ket{R} + B \langle f \ket{K^- P_u}
 \eqno(6b)$$
where $A$, $B$ and $C$ denote the amplitudes for the three
transitions (4) and describe respectively:

(A). A penguin or other decay leading to the final state  via $s \bar u$
+ singlet and $\ket{R}$. This amplitude is proportional to the
$b \rightarrow c \bar c s$ CKM matrix element.

(B). A tree decay via the state $K^- P_u$ followed only by final state
interactions which do not annihilate the the initial $u \bar u $ pair.
This amplitude is is proportional to the $b \rightarrow u \bar u s$ CKM matrix
element.

(C). A tree decay followed by a final state rescattering which goes
via the $s \bar u$ + singlet state to the final state $\ket{R}$. This amplitude
denoted by $C \sqrt {3}$ is proportional to the $b \rightarrow u \bar u s$
CKM matrix element.

The relative magnitudes and strong phases of these amplitudes are model
dependent. They are simply related to the isospin and SU(3) amplitudes
conventionally used to treat $B \rightarrow K \pi$ decays and give no new
information for these analyses\cite{PBPENG,ROSGRO}. The new ingredient
introduced by flavor-topology analyses\cite{CLOLIP12} is the inclusion of the
neutral flavor-mixed meson states $K \eta$ and $K \eta'$ modes with the same
amplitudes and same parameters. This allows the extension to the $K \eta$
and $K \eta'$ modes of any dynamical or phenomenological treatment of
$B \rightarrow K \pi$ decays without introducing additional parameters.

Substituting the relations (5b) into (6b) then gives the relations
$$ \bra{\bar K^o\pi^-} T \ket{B^-} = {{A + C} \over{\sqrt 3}}; ~ ~ ~
\tilde \Gamma(B^- \rightarrow \bar K^o\pi^-) =
{{A^2 + C^2 +2 A\cdot C} \over{3}}
\eqno(7a)$$
$$ \bra{K^- \pi^o} T \ket{B^-} = {{A + C} \over{\sqrt 6}} + {B \over{\sqrt 2}}
; ~ ~ ~ \tilde \Gamma(B^- \rightarrow K^- \pi^o) =
 {{A^2 + C^2 +2 A\cdot C} \over{6}} + {B^2\over{2}}
+ {{(A + C)\cdot B} \over{\sqrt 3}}
  \eqno(7b)$$
$$ \bra{K^-\eta} T \ket{B^-} = {B\over{\sqrt 3}} +
{{\xi (A + C)} \over{\sqrt 2}} ; ~ ~ ~ $$ $$
\tilde \Gamma(B^- \rightarrow K^- \eta) =  {B^2\over{3}} +
{{\xi^2 (A^2 + C^2 +2 A\cdot C) } \over{3}} +
{{2 \xi (A + C)\cdot B} \over{\sqrt 6}}
 \eqno(7c)$$
$$ \bra{K^-\eta'} T \ket{B^-} = {{A + C} \over{\sqrt 2}} +
{B \over{\sqrt 6}}
; ~ ~ ~
\tilde \Gamma(B^-\rightarrow K^- \eta') = {{A^2 + C^2 +2 A\cdot C} \over{2}} +
{B^2\over{6}} + {{(A + C)\cdot B} \over{\sqrt 3}}
\eqno(7d)$$

Direct CP-violation asymmetries are obtained from interference between the $A$
amplitude and the $B$ and $C$ amplitudes which have different weak phases.
$$ \tilde \Gamma(B^- \rightarrow \bar K^o\pi^-)
- \tilde \Gamma(B^+ \rightarrow K^o\pi^+) =
{{2A\cdot (C - \bar C)} \over{3}}
\eqno(8a)$$
$$ \tilde \Gamma(B^- \rightarrow K^- \pi^o) -
\tilde \Gamma(B^+ \rightarrow K^+ \pi^o) =
 {{2 A\cdot (C - \bar C)} \over{6}} +
 {{A\cdot (B - \bar B)} \over{\sqrt 3}}
  \eqno(8b)$$
$$ \tilde \Gamma(B^- \rightarrow K^- \eta) -
\tilde \Gamma(B^+ \rightarrow K^+ \eta ) =
{{2 \xi A\cdot (B - \bar B)} \over{\sqrt 6}}
 \eqno(8c)$$
$$ \tilde \Gamma(B^-\rightarrow K^- \eta') -
\tilde \Gamma(B^+\rightarrow K^+ \eta') = {{2 A\cdot (C - \bar C)} \over{2}} +
{{A\cdot (B - \bar B)} \over{\sqrt 3}}
\eqno(8d)$$
where an overall phase convention is defined in which the $A$ amplitude has the
same phase for $B^-$ and $B^+$ decays and $\bar B$ and $\bar C$ respectively
denote the $B$ and $C$ amplitudes for $B^+$ decays.

The $A$ amplitude is dominated by the penguin and expected to be much larger
than the $B$ and $C$ amplitudes. Thus $\Gamma(B^- \rightarrow K^- \eta)$ is
expected to be much smaller than for the other decays. However, to first order
in the small parameter $\xi$ and the small ratios $B/A$ and $C/A$,
  $$ {{\tilde \Gamma(B^- \rightarrow K^- \eta) -
\tilde \Gamma(B^+ \rightarrow K^+ \eta )}
\over
{\tilde \Gamma(B^- \rightarrow K^- \eta) +
\tilde \Gamma(B^+ \rightarrow K^+ \eta )}} \approx
{{3 \xi A\cdot (B - \bar B)} \over{\sqrt 6 B^2}}
 \eqno(9)$$
This is of order $(\xi A/B)$ while the analogous relative
asymmetries for other decay modes are of order $(B/A)$ and $(C/A)$. Thus even
though the signal for $CP$ violation (8c) may be small for
$B^+ \rightarrow K^+ \eta$, the signal/background ratio (9) may be more
favorable. An exact theoretical calculation of $\xi$ is not feasible. A good
estimate from future data may enable a choice  between different
decay modes as candidates for observation of direct CP violation.

Higher resonances can be incorporated into the final state rescattering with
simplifications from $C$, $P$ Bose symmetry and flavor $SU(3)$ symmetry. Since
the vector-pseudoscalar states have opposite parity, the next higher
quasi-two-body final states allowed by conservation laws are the vector-vector
s-wave and d-wave states. These can be incorporated by using models for the
decays of a scalar resonance into these channels and inputs from polarization
measurements and branching ratios for the vector-vector states.

The same approach can be used to treat vector-pseudoscalar final states.
We label corresponding quantities by subscripts $VP$ and $KV$ for
$K^*$-pseudoscalar and kaon-vector decays respectively.
Expressions for the $K\rho$ decay modes are obtained directly from eqs. (7)
and (8) for the $K\pi$ modes and the $K\omega$ and $K\phi$ decays are
giben by eqs. (1).

The $K^*P$ system differs from $KP$ because the relative phase of the
the strange and nonstrange contributions of the $\eta$
and $\eta'$ are reversed\cite{PHAWAII,PengSU3}. The analogs of eqs. (5b -8)
are thus
 $$\ket{R_{VP}} = {1\over{\sqrt 6}}\cdot \ket{K^{*-} \pi^o} -
{1\over{3 \sqrt 2}}\cdot \ket{K^{*-} \eta' } +
{2\over{3}}\cdot \ket{K^{*-} \eta } +
{1\over{\sqrt 3}}\cdot\ket{\bar K^{*o}\pi^-}
 \eqno (10)$$
$$ \bra{f} T \ket{B^-} = (A_{VP} + C_{VP} )\langle f \ket{R} +
B_{VP} \langle f \ket{K^{*-} P_u}
 \eqno(11)$$
$$ \tilde \Gamma(B^- \rightarrow \bar K^{*o}\pi^-) =
{{A_{VP}^2 + C_{VP}^2 +2 A_{VP}\cdot C_{VP}} \over{3}}
\eqno(12a)$$
$$ \tilde \Gamma(B^- \rightarrow K^{*-} \pi^o) =
 {{A_{VP}^2 + C_{VP}^2 +2 A_{VP}\cdot C_{VP}} \over{6}} + {B_{VP}^2\over{2}}
+ {{(A_{VP} + C_{VP})\cdot B_{VP}} \over{\sqrt 3}}
  \eqno(12b)$$
$$ \tilde \Gamma(B^- \rightarrow K^{*-} \eta) =  {B_{VP}^2\over{3}} +
{{4 (A_{VP}^2 + C_{VP}^2 +2 A_{VP}\cdot C) } \over{9}} +
{{4 (A_{VP} + C_{VP})\cdot B} \over{3 \sqrt 3}}
 \eqno(12c)$$
$$ \tilde \Gamma(B^-\rightarrow K^{*-} \eta') = {{A_{VP}^2 + C_{VP}^2 +
2 A_{VP}\cdot C_{VP}} \over{18}} +
{B_{VP}^2\over{6}} - {{(A_{VP} + C_{VP})\cdot B} \over{3 \sqrt 3}}
\eqno(12d)$$
$$ \tilde \Gamma(B^\pm \rightarrow K^{*\pm} \eta') + \tilde \Gamma(B^\pm
\rightarrow K^{*\pm} \eta) =  \tilde \Gamma(B^\pm \rightarrow K^{*\pm} \pi^o)
+  \tilde \Gamma(B^\pm \rightarrow \tilde K^{*o} \pi^\pm)
\eqno(13)                                          $$
$$ \tilde \Gamma(B^- \rightarrow \bar K^{*o}\pi^-)
- \tilde \Gamma(B^+ \rightarrow K^{*o}\pi^+) =
{{2A_{VP}\cdot (C_{VP} - \bar C_{VP})} \over{3}}
\eqno(14a)$$
$$ \tilde \Gamma(B^- \rightarrow K^{*-} \pi^o) -
\tilde \Gamma(B^+ \rightarrow K^{*+} \pi^o) =
 {{2 A_{VP}\cdot (C_{VP} - \bar C_{VP})} \over{6}} +
 {{A_{VP}\cdot (B_{VP} - \bar B_{VP})} \over{\sqrt 3}}
  \eqno(14b)$$
$$ \tilde \Gamma(B^- \rightarrow K^{*-} \eta) -
\tilde \Gamma(B^+ \rightarrow K^{*+} \eta ) =
{{8 A_{VP}\cdot (C_{VP} - \bar C_{VP})} \over{9}} + {{4 A_{VP}\cdot
(B_{VP} - \bar B_{VP})} \over{3\sqrt 3}}
 \eqno(14c)$$
$$ \tilde \Gamma(B^-\rightarrow K^{*-} \eta') -
\tilde \Gamma(B^+\rightarrow K^{*+} \eta') = {{2 A_{VP}\cdot
(C_{VP} - \bar C_{VP})} \over{18}}
- {{A_{VP}\cdot (B_{VP} - \bar B_{VP})} \over{3\sqrt 3}}
\eqno(14d)$$

Here $\Gamma(B^- \rightarrow K^{*-} \eta')$ is expected to be much smaller than
the other decay widths. The reversal of $\eta'/\eta$ ratio has been
suggested\cite{PHAWAII,PengSU3} as a test for the presence of an
additional component in the $\eta'$ with a quantitative prediction based only
on the $A_{VP}$ amplitude. A relation which also includes the contributions
from $B_{VP}$ and $C_{VP}$ is
$$ \tilde \Gamma(B^\pm \rightarrow K^{*\pm} \eta') =
(1/3)\cdot  \tilde \Gamma(B^\pm \rightarrow K^{*o} \pi^\pm) -
(1/3)\cdot  \tilde \Gamma(B^\pm \rightarrow K^{*\pm} \pi^o) +
{B_{VP}^2\over{3}} \eqno(15)                                          $$

We now note one other case which also suggests that A...Z allowed transitions
via
a $q \bar q $ + singlet intermediate state may be a general feature of final
state interactions which warrants further investigation.
Such a transition can produce the dramatic change in color suppression
noted\cite{PHAWAII} between $D$ and $D_s$ decays which differ only by the
flavor of a spectator quark. Since they must involve annihilation of the
spectator quark they occur only for the color-favored $D^o$ decays and
color-suppressed $D_s$ decays and not vice versa. They can thus compensate for
the color-suppression not observed in the data for the $D_s$ decays into
$K^* \bar K$ and $K^*\bar K^*$ modes relative to color-favored $\phi \pi$ and
$\phi \rho$, without affecting the definite color suppression seen in VP and VV
$D^o$ decays.

The nonstrange vector-pseudoscalar modes in both $D$ and $B$ decays already
present other puzzles\cite{PHAWAII} which surprise theorists and provide
interesting opportunities for future investigations. The role of $G$ parity has
been pointed out with reference to the $\eta \pi$, $\eta' \pi$, $\eta \rho$,
and $\eta' \rho$ for the $D_s$ decays where four channels with
different parities and $G$ parities are not mixed by strong final
state interactions\cite{PHAWAII}. The same is also true for $B$ and $B_s$
decays. For the $VP$ decays, which have a definite odd parity, there are still
two channels. One has odd G-parity like the pion and couples to $\rho \pi$; the
other has exotic even $G$ and couples to $\omega \pi$, $\eta \rho$, and $\eta'
\rho$. This even-G state does not couple to any $q \bar q$ state containing no
additional gluons. It therefore does not couple to any single meson resonances,
nor to the state produced by an annihilation diagram with no gluons emitted by
the initial state before annihilation\cite{PHAWAII}. We now note that the
coupling of the even-G state is A...Z-forbidden in the present model also for
annihilation diagrams with additional gluons present because of cancellation
between contributions from the $u  \bar u$ and $d \bar d$ components of the
$\omega$, $\eta$, and $\eta'$ wave functions, whereas the two contributions add
in the $D_s \rightarrow \rho \pi$. Here the presence of charm in the
$\eta'$ wave function may be significant because of the generally overlooked
contribution of the ``backward" weak diagram $s \rightarrow c \bar u d$.

Comparison of corresponding $D_s$ and $D$ decays into final states containing
the $\eta$ and $\eta'$ mesons have been suggested\cite{PKETA} as a means to
test for the breaking of the nonet picture by additional flavor singlet
components.

Further information which may provide important clues to this complicated
four-channel system may be obtained from angular distributions in the
$K \bar K \pi$ modes, including the $K^* \bar K$ and $K \bar K^*$ modes which
decay into $K \bar K \pi^o$. The VP $K^* \bar K$ and $K \bar K^*$ modes are
not individually eigenstates of $G$ parity and have unique p-wave angular
distributions for the vector-pseudoscalar states. The $G$-parity eigenstates
are coherent linear combinations of the two with opposite phase. They have
opposite relative parity in the $K \bar K$ system, even though
they are not produced from the same resonance. This relative parity can be
observed as constructive or destructive interference in the kinematic region
in the Dalitz plot where the two $K^*$ bands overlap. In a region where s and
p waves dominate the angular distribution of the $K \bar K$ momentum in the
$K \bar K$ center of mass system of the $K \bar K \pi$ final state relative to
the pion momentum, one $G$ eigenstate will have a $\sin^2\theta$ distribution,
the other will have a $\cos^2\theta$ distribution and interference between the
two $G$ eigenstates can show up as a forward-backward asymmetry.

A general theorem from CPT invariance\cite{PENGRHO} shows that all observed CP
asymmetries must cancel when the data are summed over all final states or over
any set of final states which are eigenstates of the strong-interaction
S-matrix. Any CP asymmetry arising in a given
channel must be canceled by an opposite CP asymmetry in some other
channels. In the case of the model described by eqs. (8), this can occur only
if there is a definite relation between the $A\cdot C$ and $A\cdot B$
interference terms. Any total CP asymmetry arising in a finite set of final
states indicates significant strong interaction rescattering between these
states and others outside the set; e.g. vector-vector or multiparticle final
states. This casts doubt on theoretical estimates of direct $CP$ violation
which do not include such rescattering.

Other attempts to estimating soft strong effects on CP violation in weak
decays\cite{Donoghue} have used Regge phenomenology with parameters
obtained from fits\cite{PDG} to total cross section data. These fits are
unfortunately highly controversial and unreliable\cite{PAQMREV}. Better fits
to the same data with completely different parameters\cite{PAQMREV,Exter} have
been obtained by using the physics input described above\cite{Precoc}. A recent
Regge phenomenology calculation\cite{Falk} for $B\rightarrow K\pi$ decays using
PDG parameters\cite{PDG} shows neither a dominant effect of order unity nor an
insignificant effect of order 1 \%. Further improvement seems unlikely. In
contrast the approach presented here uses well defined physics input subject to
experimental tests. If these tests are successful they can lead the way to a
considerable simplification in future treatments of FSI.

\acknowledgments
It is a pleasure to thank Edmond Berger, Karl Berkelman, Yuval Grossman,
Yosef Nir and J. G. Smith for helpful discussions and comments. This work
was partially supported by the German-Israeli Foundation for Scientific
Research and Development (GIF).

{
\tighten

}

\end{document}